\documentclass[aps,pre,twocolumn,groupedaddress,showpacs]{revtex4}
\pdfoutput=1
\usepackage[pdftex]{graphicx}
\usepackage{amsmath}
\usepackage{amssymb}
\usepackage{listings}
\usepackage{comment}
\newcommand {\Vect} [1] {{\bf #1}}
\newcommand {\Tens} [1] {{\bf #1}}
\newcommand {\F} {\Vect{F}}
\newcommand {\M} {\Vect{M}}
\newcommand {\be} [1]   {\begin{equation}\label{#1}}
\newcommand {\ee}       {\end{equation}}
\newcommand {\dfeq}     {\stackrel{\mbox{\scriptsize def}}{=}}
\newcommand {\om} {\boldsymbol{\omega}}
\newcommand {\ph} {\boldsymbol{\varphi}}

\def\n{\Vect{n}}
\def\e{\Vect{e}}
\def\d{\Vect{d}}
\def\D{\Vect{D}}
\def\x{\Vect{x}}
\def\E{\Vect{E}}
\newcommand {\eq} [1]   {(\ref{#1})}
\newcommand {\DS} {\displaystyle}
\def\U{U}
\def\[{\left[}
\def\]{\right]}
\def\({\left(}
\def\){\right)}

\begin{document}
\title{Vector-based model of elastic bonds for DEM simulation of solids}
\author{Vitaly A. Kuzkin, Igor E. Asonov}
\email{kuzkinva@gmail.com, Igor.asonov@gmail.com}
\affiliation{Institute for Problems in Mechanical Engineering RAS,\\ Saint Petersburg State Polytechnical University}
\date{\today}
\begin{abstract}
A new model for computer simulation of solids, composed of bonded rigid body particles, is proposed.
Vectors rigidly connected with particles are used for description of deformation of a single bond.
The expression for potential energy of the bond and corresponding expressions for forces and moments are proposed.
Formulas, connecting  parameters of the model with longitudinal, shear, bending and torsional stiffnesses of the bond, are derived.
It is shown that the model allows to describe {\it any} values of the bond stiffnesses {\it exactly}. Two different
calibration procedures depending on bond length/thickness ratio are proposed.
It is shown that parameters of model can be chosen so that under small deformations the bond is
equivalent to either Bernoulli-Euler rod or Timoshenko rod  or short cylinder connecting particles. Simple expressions, connecting  parameters of V-model with geometrical and mechanical characteristics of the bond, are derived. Computer simulation of dynamical buckling of the straight discrete rod and half-spherical shell is carried out.
\end{abstract}
\maketitle
\section{Introduction}
Discrete~(or Distinct) Element Method~(DEM)~\cite{Cundal_Strack} is widely used for computer simulation of solid and free-flowing materials. Similarly to classical molecular dynamics~\cite{Hoover_MD, Allen}, in the framework of DEM the material is represented by the set of many interacting rigid body particles~(granules). Equations of particles motion are solved numerically. In free-flowing materials the
 particles interact via contact forces, dry and viscous friction forces,  electrostatic forces etc.  Simulation of solids requires additional interparticle interactions, allowing to describe stability, elasticity, strength and other intrinsic properties that distinguish solids from free-flowing materials. In practice for simulation of granular solids particles are  connected by so-called bonds~\cite{BPM, Wang}, transmitting both forces and moments. Moments are especially important for simulation of thin structures~\cite{DEMsolutions}. The bonds can be considered either as a model of interaction between different parts of one material, represented by the particles, or a model of some additional material, connecting particles~(for example, glue~\cite{BPM} or cement~\cite{Wolff}). According to the review, presented in paper~\cite{Wang}, only several models, proposed in literature, allows to describe all possible deformations of the bond~(stretching/compression, shear, bending, and torsion).
Bonded-particle model~(BPM), proposed in paper~\cite{BPM},  is widely used for simulation of deformation and fracture of solids, in particular, rocks~\cite{Khanal, Refahi, Deng} and agglomerates~\cite{Antonyuk}. Simulation of diametrical compression of circular  particle compounds is considered in paper~\cite{Khanal}.  Compression of spherical and cubic specimens is investigated in paper~\cite{Refahi}. Fluid-rock interaction is considered in paper~\cite{Deng}. Impact of a granule with a rigid wall is considered in paper~\cite{Antonyuk}. Several drawbacks of BPM, in particular, in the case of coexistence of bending and torsion of the bond, are discussed in paper~\cite{Wang}. It is noted that the main reason for the drawbacks is incremental algorithm, used in the framework of BPM. Also is should be noted that BPM contains only two independent parameters, describing bond stiffnesses, while, in general, the bond has four independent stiffnesses~(longitudinal, shear, bending and torsional). Timoshenko rod connecting particles' centers is used as a model of a bond in paper~\cite{DEMsolutions}.
The model has clear physical meaning and is applicable for thin, long bonds under small deformations. However it has low accuracy for the description of short bonds, connecting particles' surfaces. For example, the model~\cite{DEMsolutions} is not accurate in the case of glued particles. Also the generalization of the model for the case of large  nonlinear deformations of the bond is not straightforward. Another approach, based on decomposition of relative rotation of particles, is proposed in paper~\cite{Wang}. Forces and moments are represented as functions of angles, describing relative turn of the particles.
It is shown that method~\cite{Wang} is more accurate form computational point of view than incremental procedure of BPM. Though the formalism proposed in paper~\cite{Wang} is correct from mathematical point of view, it has a drawback. It is evident from the paper that if particles rotate in the same direction and there is no relative translation, then forces and moments are equal to zero. The reason is that forces and moments, proposed in paper~\cite{Wang}, depend only on relative position and orientation of the particles, while, in general, the dependence on the orientation of the particles with respect to the bond should also be taken into account.

In the present paper forces and moments, caused by the bond, are derived from the potential energy. This approach is used in classical molecular dynamics for both material points~\cite{Hoover_MD} and rigid bodies~\cite{Allen}.
The approach for construction of potential energy of interactions between rigid bodies is proposed in paper~\cite{Price}. Initially it was applied to simulation of molecular liquids~\cite{Allen}. In papers~\cite{IvKrMoFi_MTT_2003, IvKrMo_PMM_2007} similar ideas are applied to crystalline solids. In particular, analytical description of elastic properties of graphene is carried out in paper~\cite{IvKrMo_PMM_2007}. Potentials for modeling of nonlinear interactions between rigid bodies in two and three dimensional cases are proposed, for example, in papers~\cite{BeIvKrMo_MTT_2007, Ivanova_Byzov} and~\cite{Kuzkin_DAN}. In the present paper similar ideas are used for development of simple vector-based model~(further referenced to as V-model) of elastic bonds in solids. Combination of approaches, proposed in works~\cite{IvKrMoFi_MTT_2003, Zhilin_FL} and \cite{Allen, Price}, is used. Equations describing interactions between two rigid bodies in the general case are summarized. General expression for potential energy of the bond is represented via vectors rigidly connected with bonded particles. The vectors are used for description of different types of bond's deformation. The expression for potential energy corresponding to tension/compression, shear, bending, and torsion of the bond is proposed. Forces and moments acting between particles are derived from the potential energy.
Two approaches for calibration of V-model parameters for bonds with different length/thickness ratios are presented. Simple analytical formulas connecting geometrical and elastic characteristics of the bond with parameters of V-model are derived. Main aspects of numerical implementation of the model are discussed. Two examples of computer simulations using V-model are given.

\section{Pair potential interactions between rigid bodies: the general case}
Let us consider the approach for description of pair potential interactions between rigid bodies in the general case~\cite{Allen, Price, IvKrMo_PMM_2007, Zhilin_FL}. In literature the formalism is referenced to as moment interactions~\cite{IvKrMo_PMM_2007, Kuzkin_DAN}. In the present paper moment interactions are applied for description of elastic bonds between particles in solids.

Consider a system consisting of two interacting rigid body particles, marked by indexes~$i$ and $j$. In the general case particles interact via forces and moments depending on their relative position, relative  orientation, and orientation with respect to the vector connecting the particles. Let us introduce the following designations: $\F_{ij}$, $\M_{ij}$ are force and moment acting on particle~$i$ from particle~$j$. Moment~$\M_{ij}$ is calculated with respect to center of mass of particle~$i$.
In paper~\cite{IvKrMo_PMM_2007} it is shown that~$\F_{ij}$, $\M_{ij}$ satisfy Newton's Third law, its analog for moments,
 and equation of energy balance:
\be{Third NL}
\begin{array}{l}
  \DS
  \F_{ij}=-\F_{ji}, \quad \M_{ij} + \M_{ji}-\Vect{r}_{ij} \times \F_{ij} = 0,
  \\[4mm]
  \DS
  \dot{U}_{ij}= \F_{ij}\cdot\dot{\Vect{r}}_{ij} - \M_{ij}\cdot\om_i - \M_{ji} \cdot\om_j,
 \end{array}
\ee
where~$\Vect{r}_{ij} \dfeq \Vect{r}_j-\Vect{r}_i$; $\Vect{r}_i, \Vect{r}_j$ are radius vectors of particles~$i$ and~$j$; $\om_i, \om_j$ are angular velocities; ${U}_{ij}$~is internal energy of the system.

Assume that interactions between particles are potential and internal energy~$U_{ij}$ depends on particles' relative position, relative orientation, and orientation with respect to~$\Vect{r}_{ij}$. Relative position of the particles can be described by vector~$\Vect{r}_{ij}$. Therefore $U_{ij}$ should be a function of~$\Vect{r}_{ij}$. In order to introduce the dependence of~$U_{ij}$ on particles' orientation the approach, initially proposed for liquids in paper~\cite{Price} and applied for solids in paper~\cite{Kuzkin_DAN}, is used. Let us describe the orientation of particle~$i$  via the set of vectors~$\{ \n_i^k \}_{k \in \Lambda_i}$, rigidly connected with the particle, where $\Lambda_i$  is a set of indexes. Hereinafter lower
index corresponds to particle's number, upper index corresponds to vector's number.
Maximum amount of vectors is not limited and does not influence the general considerations.
Since orientations of the particles are determined by vectors~$\{ \n_i^k \}_{k \in \Lambda_i}$, $\{\n_j^m \}_{m \in \Lambda_j}$,
it follows that  internal energy has form
\be{energy}
 U_{ij} = U(\Vect{r}_{ij}, \{ \n_i^k \}_{k \in \Lambda_i}, \{\n_j^m \}_{m \in \Lambda_j}).
\ee
Let us derive the relation between forces, moments and potential energy~$U_{ij}$. Substituting  formula~\eq{energy} into equation of energy balance~\eq{Third NL} and assuming that forces~$\F_{ij}$ and moments~$\M_{ij}$ are independent on linear and angular velocities of the particles,
one can show that
\be{force def}
 \begin{array}{l}
   \DS  \F_{ij} = -\F_{ji} = \frac{\partial U}{\partial \Vect{r}_{ij}}, \quad \M_{ij} = \sum_{k \in \Lambda_i} \frac{\partial U}{\partial
   \Vect{n}_i^k}\times\Vect{n}_i^k,
   \\[4mm]
   \DS \M_{ji} = \sum_{m \in \Lambda_j} \frac{\partial U}{\partial    \Vect{n}_j^m}\times\Vect{n}_j^m.
 \end{array}
 \ee
If internal energy~\eq{energy} is known, then forces and moments are calculated using formulas~\eq{force def}.
Note that function~$U$ must satisfy material objectivity principle, i.e. must be invariant
with respect to rigid body rotation. If objectivity principle is satisfied, then forces and moments, calculated using formulas~\eq{force def}, satisfy Newton's Third law for moments automatically.
Therefore $U$ must be a function of some invariant arguments.
For instance, the following invariant values can be used: $r_{ij}, \e_{ij} \cdot \n_{i}^{k}, \e_{ji} \cdot \n_{j}^{m}, \n_{i}^{k} \cdot \n_{j}^{m}$,
$|\e_{ij} \times \n_{i}^{k}|, |\n_{i}^{k}\times\n_{j}^{m}|$, etc.,
where~$\e_{ij} \dfeq \Vect{r}_{ij}/r_{ij}, k\in\Lambda_i, m\in\Lambda_j$.
In practice the first four expressions from the list are sufficient as the
remaining invariants can be represented via their combination. These expressions
 have simple geometrical meaning. The first one is a distance
between the particles. The second and the third invariants~($\e_{ij} \cdot \n_{i}^{k}$ and $\e_{ji} \cdot \n_{j}^{m}$) describe orientation of particles~$i$ and~$j$ with respect
 to vector~$\Vect{r}_{ij}$. The fourth invariants~$\n_{i}^{k} \cdot \n_{j}^{m}$ describe
 relative orientation of the particle.
 Thus in the general case the potential of interaction between rigid bodies is represented in the following form
%
\begin{multline}
\DS
U_{ij}=U(r_{ij}, \{\e_{ij} \cdot \n_{i}^{k}\}_{k\in\Lambda_i}, \{\e_{ji} \cdot \n_{j}^{m}\}_{m\in\Lambda_j},
\\
 \{\n_{i}^{k} \cdot \n_{j}^{m}\}_{k\in\Lambda_i, m\in\Lambda_j}).
\end{multline}
%
In general, sets~$\Lambda_i, \Lambda_j$ may contain any number of vectors. However from  computational point of view it is reasonable to minimize this number.

\section{Vector-based model of a single bond}
\begin{figure}[!ht]
\centering
\includegraphics[scale = 0.17]{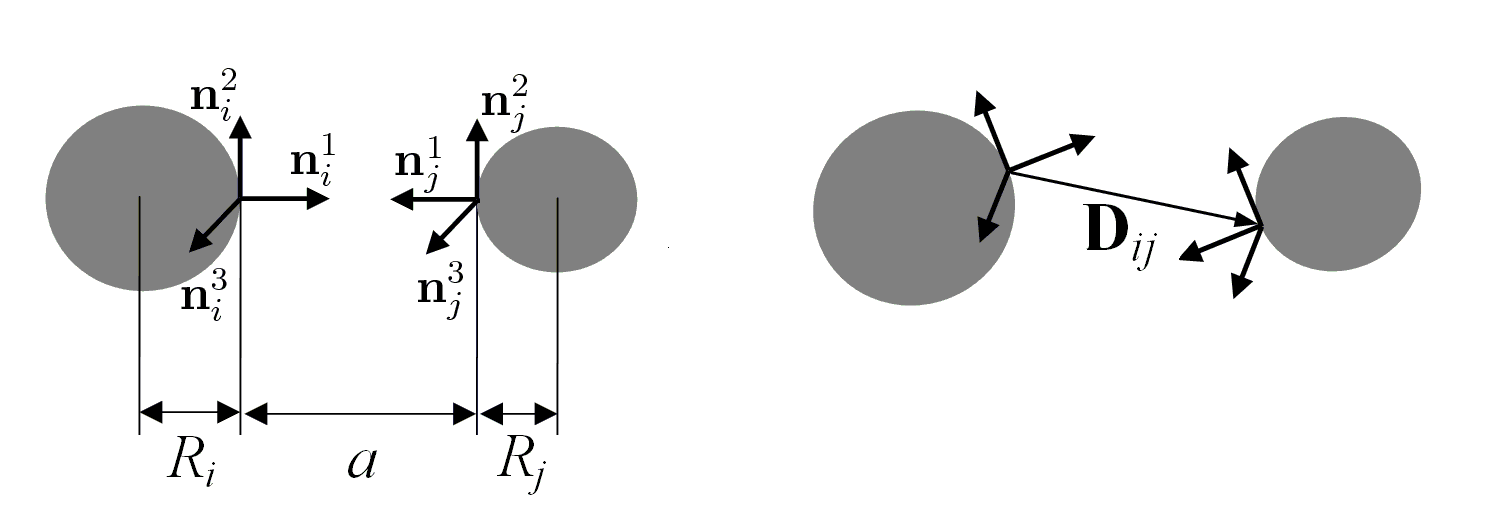}
\caption{Two bonded particles in the undeformed  state (left) and deformed state (right). Here and below $a$ is an equilibrium distance.}
 \label{materials}
\end{figure}

\begin{figure}[!ht]
\centering
\includegraphics[scale = 0.125]{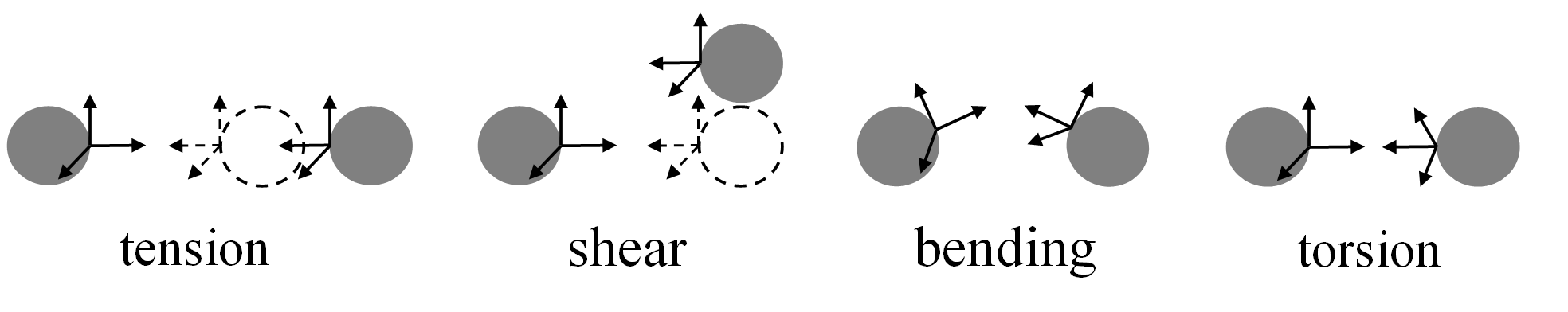}%
\caption{Different kinds of deformation of the bond and corresponding change in vectors, connected with the particles. Dashed lines show initial state of the particles.}
\label{bond_def}
\end{figure}

Let us use moment interactions for description of elastic deformation of the bond. Note that, in general, the particle can be bonded with any number of neighbors. However the behavior of the bonds is assumed to be independent. Therefore for simplicity only two bonded particles~$i$ and $j$ are considered. Assume that the bond connects two points that belong to the particles. The points lie on the line connecting the particles' centers in the initial~(undeformed) state. For example, the points can coincide with particles centers. Let us denote distance from the points to particles' centers of mass as~$R_i$, $R_j$ respectively~(see figure~\ref{materials}). For example, in the case, shown in figure~\ref{materials}, the points lie on particles' surfaces and values~$R_i$, $R_j$ coincide with particles' radii. Let us introduce  orthogonal unit vectors~$\n_i^1, \n_i^2, \n_i^3$ and~$\n_j^1, \n_j^2, \n_j^3$, rigidly connected with particles~$i$ and $j$ respectively. Lower indexes correspond to particles' numbers, upper indexes correspond to vectors' numbers.
Assume that in the undeformed state the following relations are satisfied:
\be{init}
\DS
 \n_i^1=-\n_j^1=\e_{ij}, \qquad \n_i^2=\n_j^2, \qquad \n_i^3=\n_j^3.
\ee
%
Following the idea, described in the previous paragraph, let us represent the potential energy of the bond as a function of vector~$\Vect{D}_{ij} \dfeq \Vect{r}_{ij} + R_j\n_j^1 - R_i\n_i^1$ and vectors~$\n_i^k, \n_j^m, k,m=1,2,3$. Vector~$\Vect{D}_{ij}$ connects the ``bonded''  points with radius vectors~$\Vect{r}_i + R_i \n_i^1, \Vect{r}_j + R_j \n_j^1$~(see figure~\ref{materials}).
Let us consider the following form for potential energy of the bond~$U$:
\be{general-V-model}
\begin{array}{l}
 \DS U = U_L\(D_{ij}\) + U_B(\n_i^1\cdot \n_j^1, \d_{ij} \cdot \n_i^1, \d_{ji} \cdot \n_j^1) + \\[2mm]
 \DS \qquad\qquad {} + U_T\(\{\n_i^k\cdot \n_j^k, \d_{ij}\cdot\n_i^k, \d_{ji}\cdot\n_j^k\}_{k=2,3}\), \\[4mm]
\DS D_{ij} = |\Vect{D}_{ij}|,  \quad  \d_{ij} =\Vect{D}_{ij}/ D_{ij}.
 \end{array}
\ee
Note that potential energy~\eq{general-V-model} satisfies  objectivity principle, i.e it is invariant with respect to rotation of the system as a rigid body. Let us describe the relation between functions~$U_L, U_B, U_T$ and different kinds of deformation of the bond, shown in figure~\ref{bond_def}. Function~$U_L$ describes stretching/compression, function~$U_B$ describes bending and shear of the bond. Arguments~$\d_{ij} \cdot \n_i^1, \d_{ji} \cdot \n_j^1$ change in the case of bending and shear. Argument~$\n_i^1\cdot\n_j^1$ changes only in the case of bending and is invariant with respect to shear.
Function~$U_T$ changes in the case of both torsion and bending.
%
The following expressions for functions~$U_L, U_B, U_T$ from formula~\eq{general-V-model} are proposed in the present paper:
\be{V-model}
\begin{array}{l}
\DS U_L(s) = \frac{{{B_1}}}{2}{(s - a)^2},\\[4mm]
\DS U_B(s_1, s_2, s_3) = -\frac{{{B_2}}}{2}{s_1^2} - \frac{{{B_3}}}{2}\left( s_2^2 + s_3^2 \right), \\[4mm]
\DS \U_T(\{s_{1k},s_{2k},s_{3k}\}_{k=2,3}) = -\frac{B_4}{4} \cdot \\[2mm]
\DS \qquad \qquad \cdot \sum_{k=2,3}(s_{1k} + s_{2k}s_{3k})^2(1+s_{2k}^2)(1+s_{3k}^2),
\end{array}
\ee
where $a$ is an equilibrium length of
the bond~(see figure~\ref{materials}); $B_m, m=1,..,4$, are parameters of the model. Functions~\eq{V-model} are the simplest with independent longitudinal, shear, bending, and torsional stiffnesses~(see paragraph~\ref{stiffnesses}). Note that the number of parameters of V-model is equal to the number of bond stiffnesses. Further it is shown that the behavior of the bond under small deformations can be described exactly by fitting parameters of the model. For brittle materials, such as rocks~\cite{BPM}, it is sufficient as critical deformations are usually small. On the other hand it is shown below that V-model has reasonable behavior at finite deformations~(see paragraph~\ref{Examples}). Thus very flexible structures can be considered as well.
Also V-model can be generalized for nonlinear case, changing expressions for~$U_L, U_B, U_T$ and introducing new parameters into the potential. The generalization can be important, in particular, for simulation of polymer bonds~\cite{Wolff}. Note that analogous generalization of existing models, such as BPM~\cite{BPM}, is not so straightforward.

Consider formulas~\eq{V-model}. While expressions for~$U_L$ and $U_B$ are relatively simple, the expression for~$U_T$ is not. Let us describe the idea, underlining function~$U_T$, in more details. Hereinafter denote~$\widetilde{\n}_i^k \dfeq \n_i^k - \d_{ij}\d_{ij}\cdot\n_i^k$. Vectors~$\widetilde{\n}_i^k$ lie in the plane, orthogonal to the bond. Evidently the values~$\widetilde{\n}_i^k\cdot\widetilde{\n}_j^k/|\widetilde{\n}_i^k||\widetilde{\n}_j^k|, k=2,3$ change only in the case of torsion, i.e. rotation around~$\d_{ij}$.
Therefore the potential energy~$U_T$, describing torsion of the bond, can be represented in the form~$U_T(\{\widetilde{\n}_i^k\cdot\widetilde{\n}_j^k/|\widetilde{\n}_i^k||\widetilde{\n}_j^k|\}_{k=2,3})$. However this expression contains singularity in the case~$|\widetilde{\n}_i^k| = 0$ or $|\widetilde{\n}_j^k| = 0$. Though the singularity corresponds to very large deformations of the bond, it is still not desired.
In order to avoid the singularity the following arguments of function~$U_T$ are used~(see formula~\eq{V-model} for~$U_T$)
\be{UTarg}
\begin{array}{l} \DS \frac{\(\widetilde{\n}_i^k\cdot\widetilde{\n}_j^k\)^2}{|\widetilde{\n}_i^k|^2|\widetilde{\n}_j^k|^2} (1 - (\d_{ij}\cdot\n_i^k)^4)(1 - (\d_{ji}\cdot\n_j^k)^4) = \\ [4mm]
 \quad = \(\n_i^k\cdot\n_j^k + \d_{ij}\cdot\n_i^k\ \d_{ji}\cdot\n_j^k\) \cdot \\[2mm] \qquad \cdot (1 + (\d_{ij}\cdot\n_i^k)^2)(1 + (\d_{ji}\cdot\n_j^k)^2) ,  \quad k=2,3.
\end{array}
\ee
In general, expressions~\eq{UTarg} are not invariant with respect to bending as well as~$U_T$, given by formula~\eq{V-model}. However further it is shown that in the case of small deformations~$U_T$ does not contribute to bending stiffness~(see formula~\eq{CACDCB}).

Using formulas~\eq{force def} and~\eq{V-model}, one can obtain the following formulas for~$\F_{ij}$ and~$\M_{ij}$:
\be{eq10}
\begin{array}{l}
\DS    \F_{ij} =
B_1 \( D_{ij} - a\) \Vect{d}_{ij} -
\frac{B_3}{D_{ij}} \d_{ij} \cdot\( \Vect{n}_i^1 \widetilde{\n}_i^1 +  \Vect{n}_j^1\widetilde{\n}_j^1\) + \\[2mm]
\DS \qquad \qquad \qquad  {} + \frac{1}{D_{ij}} \sum_{k=2,3} \(\frac{\partial U_T}{\partial s_{2k}}\widetilde{\n}_i^k - \frac{\partial U_T}{\partial s_{3k}}\widetilde{\n}_j^k\),
\\[6mm]
\DS  \M_{ij} = R_i \n_i^1\!\times\!\F_{ij}\! - \!\Vect{n}_i^1\!\cdot\!\[B_2 \Vect{n}_j^1\Vect{n}_j^1
 + B_3\d_{ij}\d_{ij}\]\!\times\!\Vect{n}_i^1  + \\[2mm]
 \DS \qquad \qquad \qquad  {} + \sum_{k=2,3} \(\frac{\partial U_T}{\partial s_{1k}}\n_j^k + 
\frac{\partial U_T}{\partial s_{2k}}\d_{ij}\)\!\times\!\n_i^k,
\end{array}
\ee
where~$\widetilde{\n}_i^k = {\n}_i^k-{\n}_i^k\cdot\d_{ij}\d_{ij}$. The expressions
for partial derivatives~$\partial U_T/\partial s_{mk}, m=1,2,3, k=2,3$ are the following:
\be{UTsm}
\begin{array}{l}
\DS \frac{\partial U}{\partial s_{1k}} = -\frac{B_4}{2} (s_{1k} + s_{2k}s_{3k})(1+s_{2k}^2)(1+s_{3k}^2), \\[4mm]
\DS \frac{\partial U}{\partial s_{2k}} = -\frac{B_4}{2} (s_{1k} + s_{2k}s_{3k})(1+s_{3k}^2) \cdot \\[2mm]
\DS \qquad \qquad \qquad {} \cdot ( s_{3k} + s_{1k}s_{2k} + 2 s_{3k}s_{2k}^2),\\[4mm]
\DS  \frac{\partial U}{\partial s_{3k}} = -\frac{B_4}{2} (s_{1k} + s_{2k}s_{3k})(1+s_{2k}^2) \cdot \\[2mm]
\DS \qquad \qquad \qquad {} \cdot ( s_{2k} + s_{1k}s_{3k} + 2 s_{2k}s_{3k}^2),  \ k=2,3.
\end{array}
\ee
Thus formulas~\eq{eq10}, \eq{UTsm} are used for calculation of forces and moments, acting on the bonded particles. Note that in contrast to incremental procedure~\cite{BPM}, V-model allows to calculate forces and moments at every moment of time~(time step) independently.
\\

\section{Parameters calibration}\label{calibration}
\subsection{Bond stiffnesses}\label{stiffnesses}
Let us choose parameters of V-model~$B_m, m=1,..,4$ in order to describe elastic properties of the bond  in the case of small deformations exactly. Following the idea, proposed
 in paper~\cite{IvKrMo_PMM_2007}, let us introduce stiffnesses of the bond.
 Consider the force~$\F_{ij}$ and moment
\be{M}
 \M \dfeq \M_{ij} - \(R_i\n_i^1 + \Vect{D}_{ij}/2\)\times \F_{ij},
 \ee
calculated with respect to the center of the bond, defined by vector~$\Vect{r}_i+R_i\n_i^1 + \D_{ij}/2$. According to the results of paper~\cite{IvKrMo_PMM_2007},
 under small deformations~$\F_{ij}$ and~$\M$ can be represented in the following form
\be{FijM}
\begin{array}{l}
  \DS \F_{ij} \!=\!\Tens{A}\! \cdot \! \(\!\Vect{u}_j \!-\! \Vect{u}_i\! -\! (\!R_i \ph_i \!+\! R_j\ph_j\!)\! \times \! \d_{ij}\!
  +\! \frac{1}{2} \D_{ij}\!\times\!(\!\ph_i\! +\! \ph_j\!)\!\), \\[4mm]
  \DS \M = \Tens{G} \cdot (\ph_j - \ph_i),
  \end{array}
\ee
where $\Tens{A}$, $\Tens{G}$ are stiffness tensors;  $\Vect{u}_i$, $\ph_i$ are displacement and
vector of small turn of particle~$i$. In the case of transversally symmetrical bonds,
considered in the present paper, the stiffness tensors have form
\be{stiff_tens}
\begin{array}{l}
  \Tens{A} = c_A \d_{ij}\d_{ij} + c_D (\E - \d_{ij}\d_{ij}), \\[4mm] \Tens{G} = c_B (\E - \d_{ij}\d_{ij}) + c_T \d_{ij}\d_{ij},
\end{array}
\ee
where~$\Tens{E}$ is a unit tensor. The values~$c_A, c_D, c_B, c_T$ are further referenced to as longitudinal, shear, bending, and torsional stiffness respectively. One can see from formulas~\eq{FijM}, \eq{stiff_tens} that the stiffnesses completely determine the behavior of the bond in the case of small deformations.

Let us derive the relations  between parameters of potential~\eq{V-model} and bond stiffnesses.
 First consider the expression~\eq{eq10} for force~$\F_{ij}$ in the case of pure tension:
\be{}
 \F_{ij} = B_1 \(D_{ij} -  a\) \e_{ij} =  B_1 \(|r_{ij}-R_i-R_j| -  a\) \e_{ij} .
\ee
Therefore according to formula~\eq{FijM} longitudinal stiffness of the bond~$c_A$ is equal to~$B_1$. Let us determine the relation between shear stiffness~$c_D$ and parameter~$B_3$. Consider the following deformation of the bond.
 Assume that position of particle~$i$ is fixed and particle~$j$ has a displacement~$u_j \Vect{k}$, where~$\Vect{k}$ is orthogonal to the line connecting particles in the undeformed state. Orientations of both particles are fixed.
 In this case the first formula from~\eq{FijM} has form
\be{Fshear}
    \F_{ij} \cdot \Vect{k} = c_D u_j.
\ee
Let us expand the expression~\eq{eq10} for~$\F_{ij}$ into series, assuming that~$|u_j/a| \ll 1$  and neglecting the second order terms. In this case the projection of~$\F_{ij}$ on vector~$\Vect{k}$ has form~\eq{Fshear}.
Omitting the derivation let us present the final expression for~$c_D$:
\be{}
  \quad c_D = \frac{2B_3}{a^2}.
\ee

Let us obtain analogous relation for bending stiffness of the bond~$c_B$. Assume that vector~$\D_{ij}$ remains fixed in the equilibrium state, while the particles are rotated by vectors of small turn~$\ph_i, \ph_j$.
In this case vectors~$\n_i^k, \n_j^m$ in the current~(deformed) configuration can be calculated as follows
\be{Pois}
\begin{array}{l}
\n_i^k \approx \n_i^k(0) + \ph_i\times \n_i^k(0), \\[3mm] \n_j^k \approx \n_j^k(0)+ \ph_j\times \n_j^k(0),\quad k=1,2,3.
\end{array}
\ee
Here zero denotes initial configuration, for example,~$\n_i^1(0)=-\n_j^1(0)=\e_{ij}(0)$.
This deformation corresponds to bending of the bond.
Substituting~\eq{eq10}, \eq{Pois} into~\eq{M} and leaving the first order terms only,
one obtains:
\be{Mij}
  \M \approx \[\(\frac{B_3}{2}+B_2\) \(\Tens{E} - \d_{ij}\d_{ij}\) + B_4\d_{ij}\d_{ij} \]\!\cdot\!(\ph_j-\ph_i),
\ee
The expressions for bending stiffness~${c_B}$ and torsional stiffness~$c_T$ follows from the comparison of formula~\eq{Mij} with the second formula from~\eq{FijM}. As a result the expressions relating parameters of V-model to bond stiffnesses have form
\be{CACDCB}
\begin{array}{l}
\DS  c_A = B_1,  \quad
{c_D} =
\frac{{2{B_3}}}{a^2},
\quad
{c_B} = \frac{B_3}{2} + B_2, \quad c_T = B_4.
\end{array}
\ee
It follows from formulas~\eq{CACDCB} that
choosing parameters~$B_m, m=1,..,4$ one can fit any values of the
stiffnesses. Therefore linear elastic behavior of the bond can be described exactly.
Note that no assumptions about bond's length/thickness ratio are made.

Thus if stiffnesses of the bond are known, then calculation of V-model parameters is straightforward. In principle, the stiffnesses can be measured, performing the experiments on tension, shear, bending, and torsion for the system of two bonded particles. In this case formulas~\eq{CACDCB} are sufficient for calibration. However if the body, for example, agglomerate, contains many bonds with different geometrical characteristics, then experimental calibration is practically impossible. Therefore additional model connecting the stiffnesses with geometrical and physical characteristics of the bond, such as bond length, shape, cross section area, elastic moduli of bonding material, etc., is required. Evidently the behavior of
the bond strongly depends on bond's length/thickness ratio. Therefore models used for calculation of the stiffnesses should be different for the different
ratios. Two procedures for long and short bonds are proposed below.

\subsection{Calibration for long bonds: Bernoulli-Euler and Timoshenko rod theories}
Assume that bonds are relatively long~(length/thickness ratio is larger than unity). In this case elastic rod, connecting particles, can be used a model of the bond. Comparison of V-model with the results of Bernoulli-Euler and Timoshemko rod theories~\cite{Zhilin_rod} is used as a theoretical basis for calibration. Note that in contrast to paper~\cite{DEMsolutions}, in the framework of V-model the bonds, connecting, for example, particle surfaces can be considered. This fact is important for simulation of solids, composed of glued particles, for example,  ceramic-polymer composites~\cite{Wolff}.

Let us derive the relation between parameters of  V-model and massless Bernoulli-Euler
rod connecting particles~(the rod connects points with radius-vectors~$\Vect{r}_i + R_i \n_i^1$ and $\Vect{r}_j + R_j \n_j^1$). Assume that the rod has equilibrium length~$a$, constant cross section, and isotropic bending stiffness. The expressions for longitudinal, shear, bending, and torsional stiffnesses of Bernoulli-Euler
rod are derived in paper~\cite{Berinskiy_NTV_2010}:
\be{BE_stiffnesses}
c_A = \frac{{EA}}{a},\quad {c_D} =
\frac{{12EJ}}{{{a^3}}},\quad {c_B} = \frac{{EJ}}{a}, \quad c_T = \frac{G J_p}{a},
\ee
where $E, G, A, J, J_p$ are Young's modulus, shear modulus, cross section area, moment of inertia,  and polar moment of
inertia of the cross section respectively.
For example, for the rod with circular cross section~
\be{}
    J=\frac{\pi d_b^4}{64}, \qquad  J_p = 2 J, \qquad A = \frac{\pi d_b^2}{4},
\ee
where $d_b$ is a diameter of the rod. Using formulas~\eq{CACDCB} and~\eq{BE_stiffnesses}
one obtains the expressions, connecting parameters of V-model
with characteristics of the rod
\be{BE}
{B_1} = \frac{{EA}}{a},\quad {B_2} =
-\frac{{2EJ}}{a},\quad {B_3} = - 3B_2, \quad B_4 =  \frac{G J_p}{a}.
\ee
Formula~\eq{BE} can be used for calibration of the  parameters in the case of long bonds.
If the parameters are determined by formula~\eq{BE}, then under small deformations V-model is equivalent to Bernoulli-Euler rod connecting particles. Note that in this case values~$\tilde{B}_m \dfeq B_m a, m=1,..,4$, do not depend on the equilibrium bond length~$a$. Therefore $\tilde{B}_m$ are the same for bonds with different length, but equal cross section and elastic properties. Using this fact one can reduce the number of parameters, stored in RAM, in computer simulation of systems with bonds of different length.

Bernoulli-Euler model provides simple theoretical basis for calibration.
However if length and thickness of the bond are comparable, then this model is no longer applicable~\cite{Zhilin_rod}. In this case more accurate models are required. Calibration using  Timoshenko model~\cite{Zhilin_rod} is described below.

Consider Timoshenko rod of length~$a$ and constant cross section with spherical inertia tensor. Let us derive the expressions, connecting parameters of the rod with its stiffnesses. Longitudinal and torsional stiffnesses are determined by formulas~\eq{BE_stiffnesses}. Without loss of generality the derivation of expressions for shear and bending stiffnesses is carried out in two dimensional case. Consider pure shear of the rod.  Corresponding
 system of equilibrium equations and boundary conditions for the rod has form~\cite{Zhilin_rod}:
\be{T_shear}
\begin{array}{l}
\DS w''(s) = \theta'(s), \quad  \theta''(s) + \frac{\kappa A}{2J(1+\nu)}(w'(s) -\theta(s)) = 0, 
\end{array}
\ee
\be{BC_shear}
\DS w(0) = 0, \quad \theta(0) = 0, \quad w(a) = u_j, \quad \theta(a) = 0,
\ee
where~$\nu$ is Poisson's ratio of material of the bond; $w(s)$ and $\theta(s)$ are deflection and angle of turn for the cross section with coordinate~$s$; $\kappa$ is dimensionless shear coefficient~\cite{Zhilin_rod}. In general shear coefficient~$\kappa$ depends on the shape of the cross section and length/thickness ratio for the rod. Usually~$\kappa$ is obtained comparing the results of rod theory with predictions of elasticity theory. Shear coefficients for rods with different cross sections are derived in paper~\cite{Hatchinson}. For example,
 the following expression is proposed for the rods with circular cross section:
\be{}
 \kappa = \frac{6(1+\nu)^2}{7+12\nu+4\nu^2}.
\ee
On the other hand~$\kappa$ can be considered as additional fitting parameter. Solving the system of partial differential equations~\eq{T_shear} with boundary conditions~\eq{BC_shear} one obtains an expression 
for magnitude of the shear force~$Q$, acting in the rod and shear stiffness:
\be{T_Q}
\begin{array}{l}
  \DS Q = \kappa G A (w' - \theta) = c_D u_j, \\[2mm]
  \DS c_D = \frac{12 \kappa A E  J}{a(\kappa A a^2 + 24J(1+\nu))}.
\end{array}
\ee
Let us consider bending of the rod under the following boundary conditions
\be{T_bending}
  w(0) = 0, \quad \theta(0) = \varphi_i, \quad w(a) = 0, \quad \theta(a) = \varphi_j.
\ee
Solving system of equations~\eq{T_shear} with boundary conditions~\eq{T_bending} and
calculating the magnitude of the moment~$M$, acting in the middle of the rod, one obtains
\be{T_c_B}
 \DS M = EJ \theta'\(\frac{a}{2}\) = \frac{EJ}{a} \(\varphi_j-\varphi_i\).
\ee
Formula~\eq{T_c_B} gives the expression for bending stiffness of the bond.
Thus the stiffnesses of Timoshenko rod has form:
\be{T_CACBCD}
\begin{array}{l}
 \DS c_A = \frac{EA}{a},\quad {c_D} =
\frac{12 \kappa A E  J}{a(\kappa A a^2 + 24J(1+\nu))},\\[4mm]
\DS {c_B} = \frac{{EJ}}{a}, \quad c_T = \frac{G J_p}{a}.
\end{array}
\ee
Finally using formulas~\eq{T_CACBCD} one obtains the relation between parameters of V-model and Timoshenko rod:
\be{T_B}
\begin{array}{l}
   \DS B_1 = \frac{EA}{a},~~B_2 = -\frac{2EJ(\kappa Aa^2 - 12J(1+\nu))}{a(\kappa A a^2 + 24J(1+\nu))},\\[4mm]
   \DS B_3 = \frac{6 \kappa A  E  J a}{\kappa A a^2 + 24J(1+\nu)},~~B_4 = \frac{G J_p}{a}.
\end{array}
\ee
Note that in the limit~$\kappa \rightarrow \infty$ formulas~\eq{T_B} exactly coincide with  analogous formulas~\eq{BE}, obtained using Bernoulli-Euler rod theory. If formula~\eq{T_B} is used for the calibration, then for small deformation V-model is equivalent to Timoshenko rod connecting particles.

\subsection{Calibration for short bonds}
\begin{figure}[!ht]
\centering
\includegraphics[scale = 0.20]{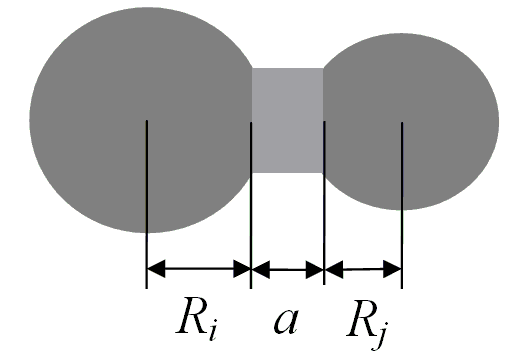}
\caption{Particles connected by a short cylinder.}
 \label{short_bond}
\end{figure}

Generally speaking the approach described above is applicable for relatively long and thin bonds with length/thickness ratio larger than unity. In the case of short bonds the models, based on elasticity theory, should be used for calibration.
Let us consider simple qualitative model, based on elasticity theory. Assume that particles are connected by a short cylinder with equilibrium length~$a$ as it is shown in
figure~\ref{short_bond}.
%
Note that in general parameters~$R_i, R_j$ are not equal to particles' radii~(the particles can even be in contact with each other). Let us derive the relations between parameters of the bond and its stiffnesses. Longitudinal stiffness~$c_A$ is, by the definition, the proportionality coefficient between force and elongation of the bond. In the case of tension  the force~$\Vect{F}_{ij}$ is created by the normal stress~$\sigma$,
acting in the bond. The following relations are satisfied:
 \be{Ft}
  \Vect{F}_{ij} \cdot \e_{ij} = \int_{(A)} \sigma dA,
 \ee
In the case of short bond, rigidly attached to the particles, the strain state of the bond is approximately uniaxial with the strain equal to~$(u_j - u_i)/a$, where~$u_i, u_j$ are particles' displacements. Then normal stress~$\sigma$ can be represented  using Hooke's law~$\sigma \approx (\lambda + 2\mu)(u_j - u_i)/a$, where~$\lambda, \mu$ are Lame coefficients for the bond. Substituting this formula into equation~\eq{Ft} one obtains
\be{}
\Vect{F}_{ij} \cdot \e_{ij} =\frac{(\lambda + 2\mu)A}{a} (u_j - u_i)  = \frac{(1 - \nu) EA}{(1 + \nu)(1 - 2\nu)a} (u_j - u_i),
\ee
Therefore longitudinal stiffness of the bond has form:
\be{ca3D}
    c_A = \frac{(1 - \nu)}{(1 + \nu)(1 - 2\nu)}\frac{EA}{a}.
\ee
One can see that longitudinal stiffness~\eq{ca3D} differs from the first formula from~\eq{T_CACBCD} by a factor of~$(1 - \nu)/((1 + \nu)(1 - 2\nu))$. Note that for nearly incompressible bonding materials the difference is crucial.

Let us derive the  expression for shear stiffness~$c_D$. Consider pure shear of the bond. Assume that position of particle~$i$ is fixed and particle~$j$ has a displacement~$u_j \Vect{k}$, where~$\Vect{k}$ is orthogonal to the line connecting particles in the undeformed state. Orientations of both particles are fixed.
In this case the force~$\F_{ij}$ is caused by shear stresses~$\tau$ acting inside the bond. Integrating the stresses over
the cross section let us represent~$\F_{ij}\cdot\Vect{k}$ in the following form
\be{shearstress}
  \F_{ij} \cdot \Vect{k}  = \int_{(A)} \tau dA,
\ee
Assume that the stress distribution over the cross section is uniform and~$\tau \approx G u_j/a$. Substituting this formula into formula~\eq{shearstress}
and comparing the result with formula~\eq{Fshear} one obtains the expression for shear stiffness:
\be{cD_short}
  c_D = \frac{GA}{a}.
\ee
One can see that the expression for shear stiffness~\eq{cD_short} and the second formula from~\eq{T_CACBCD}, derived using Timoshenko rod theory, are  qualitatively different. However it is notable that the formulas coincides in the limit of vanishing length/thickness ratio, if shear coefficient~$\kappa=1$.
Analogous  derivations for bending and torsional stiffnesses of the bond lead to the following results:
\be{cacbct}
   c_B = \frac{(1 - \nu)}{(1 + \nu)(1 - 2\nu)}\frac{EJ}{a}, \quad c_T = \frac{G J_p}{a}.
\ee
Finally using formulas~\eq{CACDCB}, \eq{cacbct} one obtains expressions, connecting the
parameters of V-model with bond characteristics:
\be{Bs}
\begin{array}{l}
\DS {B_1} = \frac{(1 - \nu)EA}{(1 + \nu)(1 - 2\nu)a},\quad {B_2} = G \[ \frac{2(1 - \nu)}{1 - 2\nu}\frac{J}{a}-\frac{Aa}{4}\],\\[4mm]
\DS {B_3} = \frac{{GAa}}{2}, \quad B_4 =  \frac{G J_p}{a}.
\end{array}
\ee
Thus in the case of short bonds formulas~\eq{Bs} can be used for calibration of V-model.

\section{On numerical implementation of V-model}
Let us describe the numerical procedure for simulation of
solids using V-model. Consider the system of~$N$ particles, connected by bonds. Other types of interactions are not considered in the present paragraph.
The system of motion equations has classical form:
\be{motion_eq}
 m_i \ddot{\Vect{r}}_i = \sum_{j\neq i}\F_{ij}, \qquad  \Theta_i \dot{\om}_i = \sum_{j\neq i}\M_{ij},
\ee
where~$m_i, \Theta_i$ are mass and moment of inertia of the particle~(for simplicity it is assumed that all particles have spherical inertia tensor). If particles~$i$ and~$j$ are bonded, then force~$\F_{ij}$ and moment~$\M_{ij}$, caused by the bond, are calculated using formulae~\eq{eq10}. Otherwise they are equal to zero. The system~\eq{motion_eq} is solved in couple with kinematic equations, connecting linear and angular velocities with positions and orientations of the particles. For example, let us determine the turn of particle~$i$ from initial orientation to current one by rotational tensor~$\Tens{P}_i$.
Then kinematic formulas are
\be{quat}
 \dot{\Vect{r}}_i = \Vect{v}_i, \qquad \dot{\Tens{P}}_i =  \om_i \times \Tens{P}_i.
 \ee
Numerical integration of equations~\eq{motion_eq}, \eq{quat} gives current positions and  orientations of the particles at every time step.

As it was discussed forces and moments between particles~$i$ and $j$ are calculated using vectors~$\n_i^k, \n_j^k, k=1,2,3$, connected with the particles. The vectors are introduced  according to formula~\eq{init} at moment~$t_{*}$, when the bond is created, and corotate with the particles. Consider the simplest approach for calculation of their current coordinates.
Let us introduce the basis, consisting of orthogonal unit vectors~$\x_i^m, m=1,2,3$, rotating with
particle~$i$. Then current orientation of vectors~$\x_i^m$ is determined as follows
\be{Eik}
  \x_i^m(t) = \Tens{P}_i(t) \cdot \x_i^m(0).
\ee
Let us use coordinates of vectors~$\n_i^k, k=1,2,3$ in the comoving basis~$\x_i^m, m=1,2,3$ for calculation of current orientation of the vectors~$\n_i^k, k=1,2,3$.
Then at each time step vectors~$\x_i^m, m=1,2,3$ are rotated using
equation~\eq{Eik} and vectors~$\n_i^k$ are determined using their coordinates~$\n_i^k\cdot\x_i^m, m,k=1,2,3$,
 stored in RAM:
\be{nik}
  \n_i^k = \sum_{m=1}^3 \(\n_i^k \cdot\x_i^m\) \x_i^m.
\ee
Note that $\n_i^k \cdot\x_i^m, k,m=1,2,3$ does not depend on time and therefore can be calculated only at $t = t_*$. The described procedure allows to avoid rotation of all vectors, connected with the particle, using equation~\eq{Eik}.

Consider calculation of forces and moments caused by the bonds. At every time step one should go over all the bonds and calculate corresponding forces and moments. Therefore
in computer code, written in object-oriented programming language, it is convenient to introduce a class ``Bond''. In general, the element of this class contains the following parameters: pointers to bonded particles, initial length of the bond~$a$, parameters~$B_m, m=1,..,4$, and coordinates of vectors~$\n_i^k, \n_j^k, k=1,2,3$ in the comoving coordinate systems. For storage of the bonds it is also convenient to introduce a class for bond list. For example, in C++ language it can be implemented using std::map.

Thus the algorithm for computer simulation using V-model is the following. At every time step:\\
1) Create new bonds if required. Calculate parameters of the bonds. Add created bonds to the list.\\
2) Check if the particles are bonded using list of the bonds. For each pair of bonded particles: get bond parameters, calculate current vectors~$\n_i^k, \n_j^k, k=1,2,3$ and length of the bond $D_{ij}$.\\
3) Calculate forces and moments between the particles using~\eq{eq10}.\\
4) Calculate linear and angular velocities  at the next time step.\\
5) Calculate positions and orientations of the particles, coordinates for vectors~$\x_i^k, k=1,2,3$ at the next time step.

\section{Examples}
\label{Examples}
In general using V-model one can  simulate mechanical behavior of any solid consisting of~(or represented by) bonded particles. However the most challenging problem for all bond models is computer simulation of one layer thin structures, such as discrete rods and shells~(see figure~\ref{rod}, \ref{shell_buckl}). In order to describe the behavior of the structures adequately bonds should transmit both forces and moments and have, generally speaking, independent longitudinal, shear, bending, and torsional stiffnesses. Therefore computer simulation of discrete rods and shells is considered below.

For simplicity assume that all particles have the same mass~$m$ and radius~$R$. The bonds connect particles' centers and have circular cross section with diameter~$d_b$. Bernoulli-Euler model is used for the calibration.
 Let us represent all values via three dimensional parameters:
equilibrium bond length~$a$\footnote{In the case of discrete shell considered below the bonds have different lengths. Thus $a$ is a length scale of the problem.}, particle mass~$m$ and longitudinal
stiffness of the bond~$c_A$. In computer code these parameters can be set equal to unity. All other parameters are represented via~$a, m, c_A$ and dimensionless values.
In particular, the  following dimensionless parameters are used:
 \be{dim}
 \begin{array}{l}
  \DS \frac{Ea}{c_A} = \frac{4}{\pi}\(\frac{a}{d_b}\)^2,\quad  \frac{A}{a^2} = \frac{\pi}{4}\(\frac{d_b}{a}\)^2, \quad \frac{J}{a^4} =\frac{\pi}{64}\(\frac{d_b}{a}\)^4, \\ [4mm]
  \DS \frac{B_1}{c_A} = 1, \quad \frac{B_2}{c_A a^2} = -\frac{1}{8}\(\frac{d_b}{a}\)^2, \quad \frac{B_3}{c_A a^2} = \frac{3}{8}\(\frac{d_b}{a}\)^2,\\ [4mm]
 \DS \frac{B_4}{c_A a^2} = \frac{1}{16(1+\nu)}\(\frac{d_b}{a}\)^2.
\end{array}
\ee
One can see that the dimensionless parameters of the bond depends only on Poisson's ratio~$\nu$ and the ratio~$d_b/a$.

\subsection{Quasistatical and  dynamical buckling of a discrete rod}
Consider initially straight discrete rod, directed along~$x$-axis and consisting of~$N$ bonded particles. Assume that the bonds connect particles' centers. First let us simulate quasistatical buckling of the rod under compression using the following procedure. Initial velocities of the particles are randomly distributed in the circle with radius~$v_0$. Initial angular velocities are set to zero.
Every~$T_{*}$ time units the uniform deformation~$\varepsilon_{*}$ is applied to the discrete rod.
After every deformation equations of particles motion~\eq{motion_eq} are integrated using leap-frog algorithm~\cite{Allen}.
 Translational degrees of freedom of the ends of the discrete rod remain fixed. The procedure is repeated until buckling.
 During the simulation compressive force acting in the rod is calculated and averaged with period~$T_{*}$.
 The following values of the parameters are used:
\be{paramrod}
\begin{array}{l}
 \DS N = 10, \quad \frac{R}{a} = 0.4, \quad \frac{\Theta}{m a^2} =  64 \cdot 10^{-3}, \quad \frac{v_0}{v_*} = 10^{-6}, \\ [4mm]
 \DS \frac{\Delta t}{T_0} = 10^{-2}, \quad  \frac{d_b}{a} = 0.2, \quad \nu = 0.2,  \quad \frac{B_1}{c_A} = 1, \\[4mm]
 \DS   \frac{B_2}{c_A a^2} = -5\cdot 10^{-3}, \quad \frac{B_3}{c_A a^2} = 15\cdot 10^{-3},
   \\[4mm]
    \DS
    \frac{B_4}{c_A a^2} = 2.08\cdot 10^{-3}, \quad \varepsilon_{*} = -10^{-7}, \quad \frac{T_*}{T_0} = 10,
\end{array}
\ee
where~$\Theta$ is particle's moment of inertia; $\Delta t$ is a time step; $T_0 = 2\pi \sqrt{m/c_A}$ is a period of small vibrations of one particle on the spring with stiffness~$c_A$; $v_* = a\sqrt{c_A/m}$ is a velocity of long waves in one-dimensional chain, composed of particles with mass~$m$, connected by springs with stiffness~$c_A$ and equilibrium length~$a$. As a result the following value of critical compressive force is obtained:~$f/(c_A a) = 3.19 \cdot 10^{-4}$. The resulting value is only~$4\%$ higher than static Euler critical force~$f_E/(c_A a) = \pi^2 EJ/(c_A a^3) = 3.05 \cdot 10^{-4}$. Note that in the framework of Bernoulli-Euler  model the critical force depends on length and bending stiffness of the rod. Therefore bending stiffness of the discrete rod, composed of particles, within~$4\%$ accuracy coincides with bending stiffness of Bernoulli-Euler rod.

\begin{figure}[!ht]
\centering
\includegraphics[scale = 0.052]{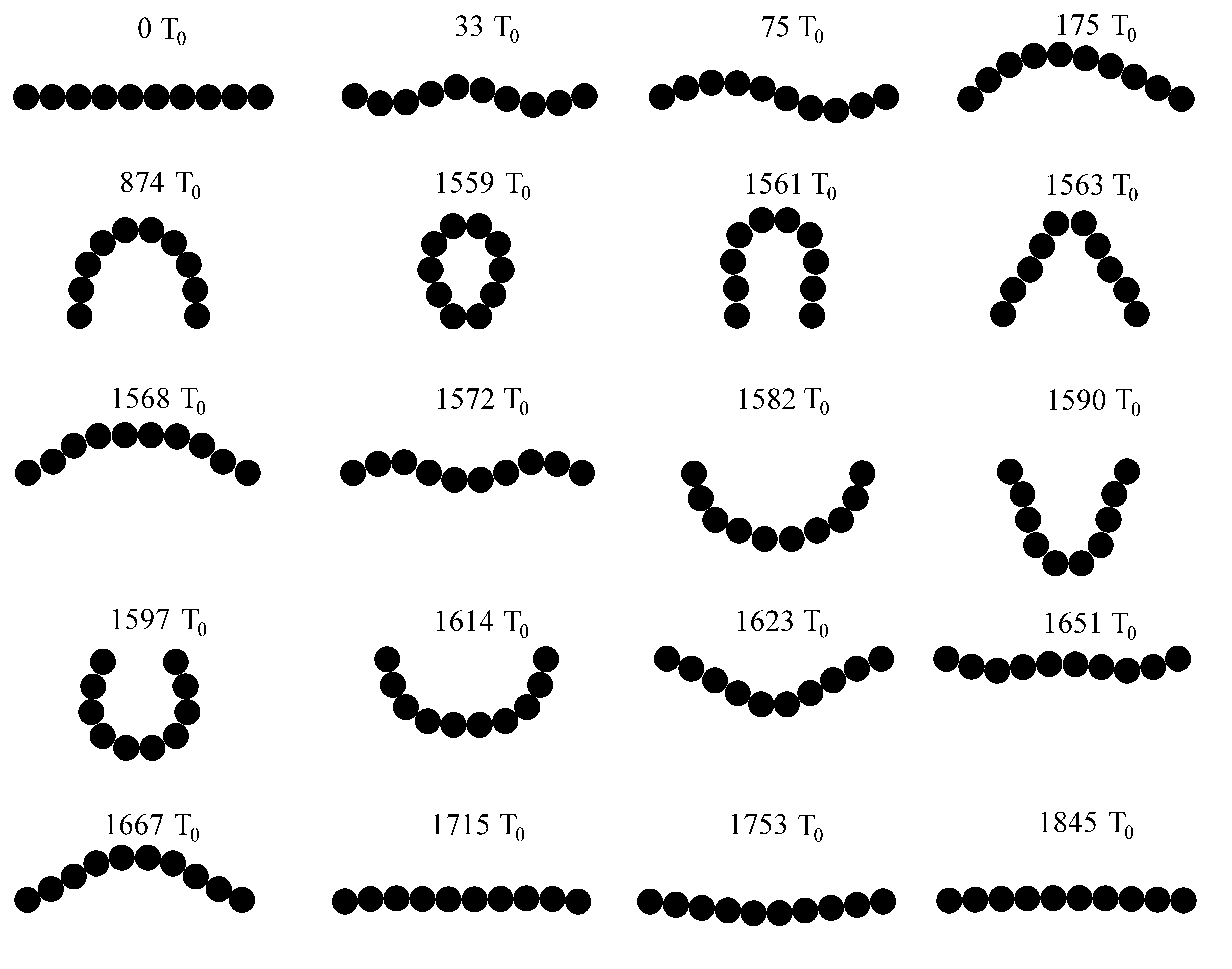}
\caption{Dynamical buckling of the discrete rod. Numbers in the figure are corresponding  moments of time. Particles radii equal 0.5a are used for visualization.}
 \label{rod}
\end{figure}

Consider dynamical buckling of the same discrete rod. In addition to V-model linear viscous forces proportional to particles velocities are introduces. Denote viscosity coefficient as~$b$. Initial velocities of the particles are randomly distributed inside the sphere with radius~$v_0$. In order to simplify visualization of the results $z$-components of the velocities for all particles are set to zero\footnote{Otherwise the buckling is performed in several planes and the visualization is not so straightforward.}. Initial angular velocities are equal to zero.
Let the ends of the rod move toward each other with constant velocities~$v_e$ until the distance between the ends becomes equal to~$a$~(see figure~\ref{rod}, $t/T_0=1559$).  Then~$x$-components of the velocities of the rod ends are released and $y-, z-$ components remain equal to zero.
The following values of dimensionless parameters are used in addition to parameters~\eq{paramrod}: $v_e/v_* = 10^{-3}, b/b_0 = 26\cdot 10^{-4}$,
where~$b_0 = 2\sqrt{m c_A}$ is a critical value of friction for two particle system.
The motion of the discrete rod is shown in figure~\ref{rod}.
%
One can see buckling and post-buckling behavior of the discrete rod. At time~$t/T_0=33$ shape of the discrete rod corresponds to the third buckling mode of Bernoulli-Euler rod. The excitation of high instability mode is typical for fast dynamical buckling. At the moment~$t/T_0=1559$ $x$-components of velocities of the rod ends are released and the rod performs strongly nonlinear free vibrations, converging to its initial straight configuration~($t/T_0 > 1845$). Therefore there is no plastic deformations.

Thus V-model allows to simulate large elastic deformations of
discrete rods including large displacements and rotations of the particles. In the case of small deformations considered above the behavior of the discrete rod is in a good agreement with Bernoulli-Euler rod theory.

\subsection{Discrete half-spherical shell under the action of point force}
\begin{figure}[!ht]
\centering
\includegraphics[scale = 0.132]{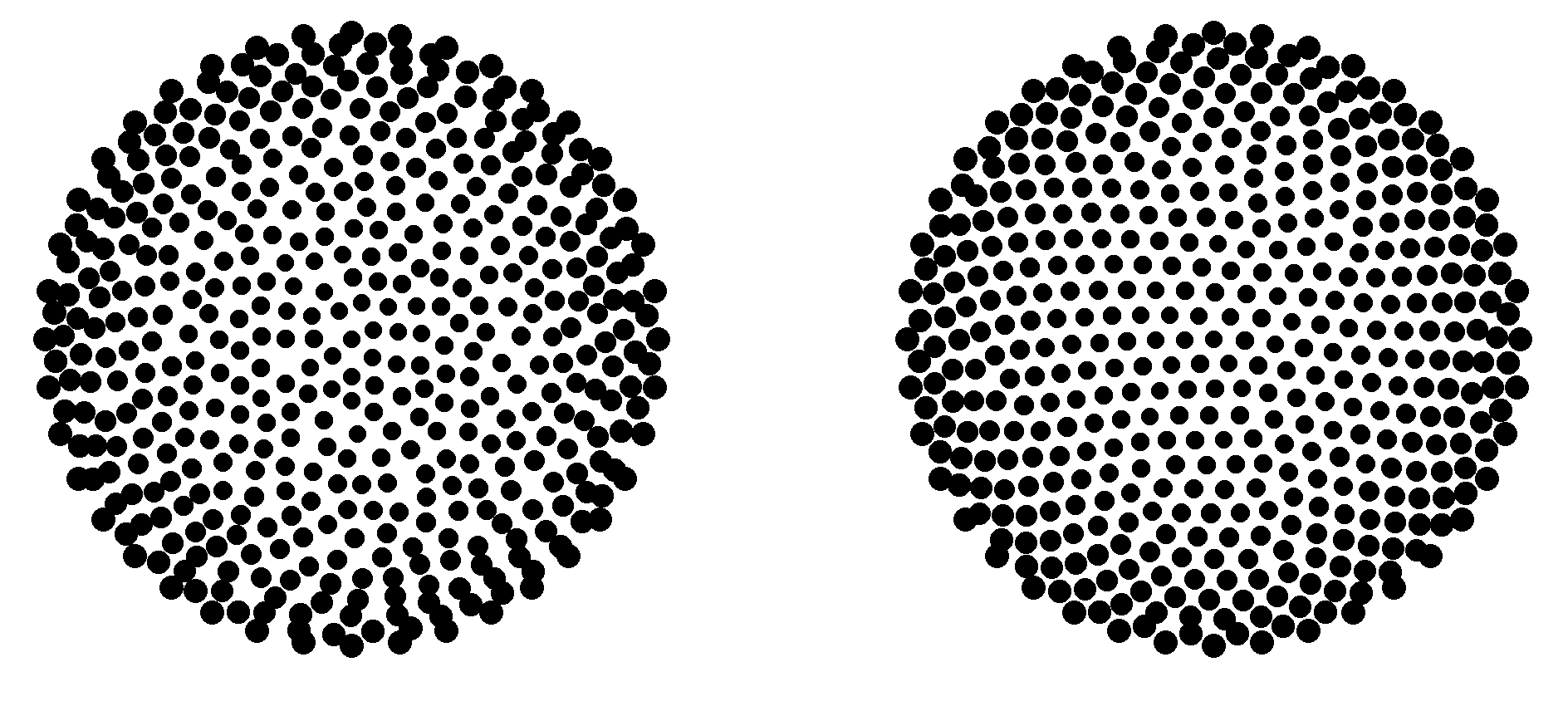}
\caption{The initial~(left) and final~(right) distributions of the particles on the half-sphere. Bottom view. Particles radii equal~$0.125 a$ are used for the visualization.}
 \label{equil}
\end{figure}

Consider dynamical buckling of discrete half-spherical shell under the action of constant point force, acting on the shell along the axis of central symmetry. First let us generate relatively uniform distribution of particles on the half-sphere. Note that this problem is identical to mesh generation problem in the framework of, for example, finite element method~(FEM). FEM packages usually use geometrical methods of mesh generation, such as triangulation. In the present paper simple particle-based method is proposed. First the circle with radius~$R_c$ of the half-sphere is created. The number of particles lying on the circle is calculated as the nearest integer value to~$2\pi R_c/a$. This particles are uniformly distributed on the circle and remain fixed during creation of the initial configuration. The other particles are generated randomly on the half-sphere. The restriction that particles can not be closer than~$0.4a$ to each other is used. Note that in this case~$a$ is a length scale of the problem. In general it is not equal to equilibrium bond length. The resulting random distribution of the particles is shown in figure~\ref{equil}~(left). Then the dynamics of translational motion of
 the particles interacting via repulsive force~$\Vect{F}_{ij}^r$ only is simulated.
 The forces are calculated according to the following formula:
\be{}
  \Vect{F}_{ij}^r = -f_0\(\frac{a}{r_{ij}}\)^8 \Vect{r}_{ij}.
\ee
The restriction~$r_i = R_c, i=1,..,N$ is applied during the simulation. The following values of the
parameters are used for the simulation:
\be{}
\begin{array}{l}
 \DS N = 458, \quad~N_s = 15 \cdot 10^3, \quad \frac{v_0}{v_*} = 0,\quad  \frac{\Delta t}{T_0} = 10^{-2}, \\[4mm]
  \DS   \frac{a_{cut}}{a} = 2.1, \quad \frac{f_0}{c_A} = 10^{-2}, \quad  \frac{b}{b_0} = 26 \cdot 10^{-5}.
\end{array}
\ee
where $a_{cut}$ is a cutoff radius; $N_s$ is a number of time steps.
The initial and final distributions of the particles are shown in figure~\ref{equil}.
%
%
One can see that resulting distribution of the particles is much more uniform than the initial one.

\begin{figure}[!ht]
\centering
\includegraphics[scale = 0.055]{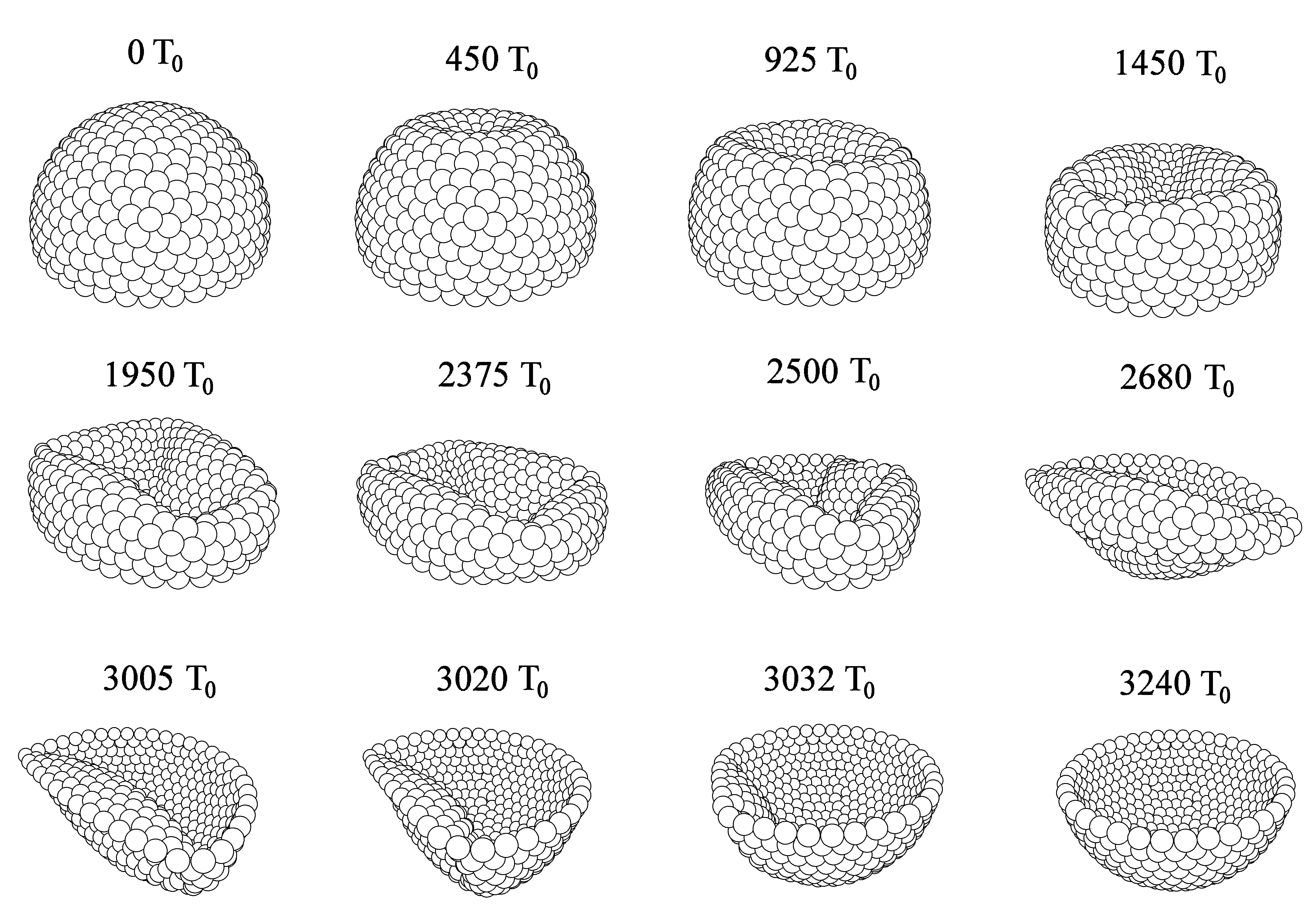}
\caption{Buckling of the discrete half-spherical shell under point force load. Particles radii equal 0.5a are used for visualization.}
 \label{shell_buckl}
\end{figure}

After creation of the initial configuration the nearest particles are bonded. For the sake of simplicity it is assumed that bonds connect particles centers. Equilibrium length for each bond is set equal to the distance between centers of the particles. Therefore there is no residual stresses in the initial state of the discrete shell. Also it is assumed that parameters of V-model~$B_m, m=1,..,4$  are the same for all bonds. Dynamical buckling of the shell under the action of constant point force of magnitude~$f_s$ is considered. The force is applied  along the axis of central symmetry of the shell until the complete  buckling. In the given example the force vanishes at~$t/T_0=3000$.  Components of displacements of the boundary particles along the symmetry axis are set to zero. In order to avoid self-penetration of the shell
 contact Hertz forces~$\Vect{F}_{ij}^H$ are introduced.
 The forces are calculated using formula
\be{}
  \DS \Vect{F}_{ij}^H =
	\left\lbrace
  	\begin{array}{lc}
  		-\frac{c_H}{\sqrt{a}}\(2R - r_{ij}\)^{\frac{3}{2}} \e_{ij}, & r_{ij} < 2R \\
  		0, & r_{ij} \geq 2R
  	\end{array}
    \right.
\ee
where~$c_H$ is a contact stiffness of the particle. Particle radius~$R$ is chosen so that~$2R$ is smaller than the minimum distance between particles in the initial configuration.
 The following values of the parameters are used for the simulation:
\be{param_shell}
\begin{array}{l}
 \DS N = 458,\quad    \frac{R}{a} = 0.35,\quad  \frac{\Theta}{m a^2} =  49 \cdot 10^{-3}, \quad \frac{v_0}{v_*} = 10^{-6},\\[4mm]
  \DS \frac{\Delta t}{T_0} = 10^{-2},\quad   \frac{b}{b_0} = 26\cdot 10^{-4}, \quad \frac{d_b}{a} = 0.2,\quad \nu = 0.2,\\[4mm]
  \DS   \frac{c_H}{c_A} = 1,\quad   \frac{f_s}{c_A a} = 10^{-2}, \quad \frac{B_1}{c_A} = 1, \quad \frac{B_2}{c_A a^2} = -5\cdot 10^{-3}, \\[4mm]
   \DS   \frac{B_3}{c_A a^2} = 15\cdot 10^{-3}, \quad
    \frac{B_4}{c_A a^2} = 2.08\cdot 10^{-3}.
\end{array}
\ee
The results of the simulation are shown in figure~\ref{shell_buckl}. Buckling and post-buckling behavior of the shell are presented.
In the places, where the shell folds, the bonds undergo extremely large turns
and deformation. For example, large deformations occur at moment~$t/T_0 = 2680$~(see figure~\ref{shell_buckl}).
However large deformations do not lead to any instability or other unphysical behavior of V-model.
%
Thus one can conclude that V-model is applicable for computer simulation of discrete shells under large displacements, turns, and deformations.

\section{Results and discussions}~\label{discussion}
In the present paper a new model for elastic bonds in solids is proposed. Vectors rigidly connected with particles are used for description of bond deformation.   The expression for potential energy of the bond as a function of the vectors is proposed.
Corresponding forces and moments acting between bonded particles are calculated using potential energy function. This approach  guarantees that the forces and moments are conservative and the bond is perfectly elastic. Dissipative terms can also be added if required. Expressions connecting  parameters of V-model with longitudinal, shear, bending, and torsional stiffnesses of the bond are derived in the case of small deformations.  It is shown that appropriate choice of the parameters allows  to describe any values of all the bond stiffnesses exactly. Two different calibration procedures depending on bond length/thickness ratio are proposed.
In the case of rod-like bonds the comparison with Bernoulli-Euler and Timoshenko rod theories is used for calibration. It is shown that parameters of V-model can be chosen so that under small deformations the bond is equivalent to either Bernoulli-Euler or Timoshenko rod connecting particles. Note that in the framework of V-model the bond may connect any two points belonging to the particles and lying on the line connecting particle centers in the initial state~(in particular, particles' centers or points lying on the surfaces). The model for calibration in the case of short bonds is proposed. In all the cases simple expressions, connecting  parameters of V-model with geometrical and mechanical characteristics of the bond, are derived. Two examples of computer simulations using V-model are given. The most challenging structures, notably one layer thin discrete rods and shells, are considered. Computer simulations of dynamical buckling of the straight discrete rod and half-spherical shell are carried out. It is shown that V-model is applicable for description of  large elastic deformations of solids composed of bonded particles.

Simulation of fracture is not considered in the present paper. However V-model allows to formulate fracture criteria for the bond. For example, the criterion, proposed in paper~\cite{BPM}, can be directly implemented in the framework of V-model.

\begin{acknowledgments}The authors are deeply grateful to Michael Wolff, Sergiy Antonyuk, Igor Berinskiy, William Hoover, and Anton Krivtsov for useful discussions and motivation for this work.
\end{acknowledgments}

\end{document}